\documentstyle[prl,aps]{revtex}
\begin{document}
\draft

\title{\bf Photon Localization and Vacuum Noise in \\ Optical Measurements}

\author{Alexander S. Shumovsky}

\address{Physics Department, Bilkent University, Bilkent, Ankara,
06533, Turkey}

\maketitle

\begin{abstract}
Description of detection and emission in terms of the photon
localization is discussed. It is shown that the standard
representation of the plane waves of photons should be revised to
take into consideration the boundary conditions caused by the
presence of quantum emitters and detectors. In turn, the change of
the boundary conditions causes spatially inhomogeneous structure
of the electromagnetic vacuum which leads to the increase of the
vacuum noise over the level predicted within the framework of the
model of plane waves of photons.
\end{abstract}


\twocolumn

Since the early days of quantum theory of light, the problem of
localizing photons has attracted a great deal of interest (e.g.,
see \cite{1,2} for recent discussion). The point is that the
photon operators of creation and destruction are defined in all
space. At the same time, the intensity measurement by means of a
photodetector with finite sensitive area $\sigma$ presupposes a
kind of the photon localization, at least in vicinity of $\sigma$
\cite{3}. The transformation of photons into an electronic signal
in photodetectors is not the only onion in the stew. Another
example of some considerable interest is provided by the emission
and absorption of radiation by atoms and molecules.

We now note that the electromagnetic field is usually quantized as
though it is free and propagates in empty space. This model leads
to well-known plane waves of photons, corresponding to the
quantized translation invariant solutions of the homogeneous
Helmholtz wave equation \cite{1,2,4}. The presence of atoms or
surfaces which are able to interact with photons leads to the
change of boundary conditions and hence to the violation of
translational symmetry. For example, the presence of a single
point-like atom causes the multipole structure of the field which
can be described in terms of quantized spherical waves \cite{5,6}.
The latter case is specified by the $SO(3)$ symmetry rather than
translational symmetry.

In this note we show that the taking account of the boundary
conditions at both emission and detection of the field from the
plane waves of photons lead to more adequate picture of the photon
localization. We also show that the change of the boundary
conditions strongly influences the zero-point oscillations of the
field strengths which causes a deterioration of quantum limit of
precision of measurements.

Consider first an atom located at the origin of the reference
frame spanned by the complex base vectors
\begin{eqnarray}
{\vec \chi}_{\pm}= \mp \frac{{\vec e}_x \pm i {\vec
e}_y}{\sqrt{2}} , \quad \quad {\vec \chi}_0={\vec e}_z. \nonumber
\end{eqnarray}
These vectors formally coincide with the states of spin $1$ of the
photon \cite{7}. Since the quantum electrodynamics defines the
spin states of photons as the polarization \cite{8}, we can choose
to interpret ${\vec \chi}_{\pm}$ as the unit vectors of
transversal polarization with either positive or negative
helicity, while ${\vec \chi}_0$ is the unit vector of linear
polarization in the $z$-direction. The third spin state is
forbidden in the case of plane waves of photons due to the
translational invariance, while allowed in the case of spherical
waves of photons \cite{5,6}. An arbitrary vector $\vec A$ is
expanded in this basis as follows
\begin{eqnarray}
{\vec A}= \sum_{\mu} (-1)^{\mu}{\vec \chi}_{- \mu}A_{\mu}.
\nonumber
\end{eqnarray}
The positive-frequency part of the electric field strength of the
monochromatic multipole field is then defined as having components
\cite{5,8}
\begin{eqnarray}
E_{\mu}({\vec r})=ik \gamma \sum_{\lambda} \sum_{j=1}^{\infty}
\sum_{m=-j}^j V_{\lambda jm \mu}({\vec r})a_{\lambda jm},
\label{1}
\end{eqnarray}
where $\lambda =E,M$ denotes the type of radiation (either
electric or magnetic), $\gamma$ is the normalization factor, $j,m$
are the angular momentum quantum numbers. In the classical
picture, the complex field amplitudes are defined by the source
\cite{4}. To obtain the quantum counterpart, we have to subject
these amplitudes to the Weyl-Heisenberg commutation relations
\cite{5}
\begin{eqnarray}
[a_{\lambda jm},a^+_{\lambda' j'm'}]= \delta_{\lambda \lambda'}
\delta_{jj'} \delta_{mm'} . \label{2}
\end{eqnarray}
The mode functions in (1) have the form \cite{5,8}
\begin{eqnarray}
V_{Ejm \mu}({\vec r}) = \frac{1}{\sqrt{2j+1}} \nonumber \\ \times
[\sqrt{j} f_{j+1} \langle 1,j+1, \mu ,m- \mu |jm \rangle Y_{j+1,m-
\mu} \nonumber \\ - \sqrt{j+1} f_{j-1} \langle 1,j-1, \mu ,m- \mu
|jm \rangle Y_{j-1,m- \mu}], \nonumber \\ V_{Mjm \mu}({\vec r})  =
f_j(kr) \langle 1,j, \mu ,m- \mu |jm \rangle Y_{j,m- \mu},
\label{3}
\end{eqnarray}
where the radial function $f_{\ell}(kr)$ is represented either by
the spherical Bessel function $j_{\ell}(kr)$, in the case of
standing waves in a spherical cavity, or by the spherical Hankel
functions of the first and the second kind, describing the
outgoing and converging spherical waves respectively. Here
$\langle \cdots |jm \rangle$ denotes the Clebsch-Gordon
coefficient of vector addition of the spin and orbital parts of
the total angular momentum and $Y_{\ell,m- \mu}$ is the spherical
harmonics.

In view of (2), the zero-point oscillations of the electric field
strength (1) have the form
\begin{eqnarray}
C_E({\vec r}) \equiv \langle 0|({\vec E}({\vec r})+{\vec
E}^+({\vec r}))^2|0 \rangle \nonumber \\ = \sum_{\mu} \langle
0|E_{\mu}({\vec r})E^+_{\mu}({\vec r})|0 \rangle  \nonumber \\ =(k
\gamma )^2 \sum_{\mu} \sum_{\lambda ,j,m} |V_{\lambda jm
\mu}({\vec r})|^2. \label{4}
\end{eqnarray}
To make a comparison, we remind here that the zero-point
oscillations in the case of the monochromatic plane waves have the
form \cite{1,2}
\begin{eqnarray}
C_{plane}= 2(k \gamma' )^2 \label{5}
\end{eqnarray}
everywhere.

It is seen that, unlike (5), the spherical waves of photons have
the spatially inhomogeneous zero-point oscillations. It is now a
straightforward matter to show that (4) strongly exceeds the
standard level, given by (5), at least in some vicinity of the
origin (atom), while tends to (5) as $kr \gg 1$. A more detailed
examination shows that $C_E({\vec r}) \gg C_{plane}$ at $kr \leq
2.5$ which gives the distance of the order of $0.3 \Lambda$ where
$\Lambda$ is the wavelength. Let us stress that this distance is
of the order of  typical interatomic separation in experiments
with trapped Ridberg atoms \cite{9}.

We now stress that the above results have been obtained under the
only assumption that the atom exists at the origin independent of
whether we use it for emission or detection. Therefore, the strong
increase of the vacuum noise in vicinity of the atom should
influence both the emission and detection processes. As a simple
model of complete Hertz-type optical experiment, we consider the
two identical atoms separated by distance $d$. The first atom
(source) is prepared initially in the excited state of some
multipole transition, while the second atom (detector) is in the
ground state. Then, the measurement consists in the emission and
successive absorption of a photon.

It is clear that in order to take into account the initial
localization of photon within the source, the radiation should be
described in terms of the outgoing spherical wave focused on the
first atom. In turn, the final localization within the detector,
assumes the converging spherical wave focused on the second atom.
To combine these two processes into the common picture, we have to
describe the filed as the superposition of outgoing and converging
waves. The coefficients of this superposed state should be defined
by the boundary conditions for the real radiation field. Taking
into account the recent investigation \cite{10}, we anticipate
that this model obeys the causality principle of the
electrodynamics. In view of the position dependence in (3) and
(4), it is clear that, at far distances ($d \gg \Lambda$), the
major contribution into the vacuum noise of measurement comes from
the detecting atom, while, at intermediate and short distances,
the noise of measurement is increased due to the influence of the
source atom.

Consider now the measurement of a plane photon by a photodetector.
At far distances, the photon is described by a unique wave vector
$\vec k$. The Mandel's localization in vicinity of the sensitive
area $\sigma$ \cite{3} assumes that the wave converges to
$\sigma$. This means that there is a variety of directions of the
wave vectors near $\sigma$, although all of them have the same
length. This picture, based on the taking account of the boundary
conditions, can be described by a proper expansion over spherical
waves. In view of the above discussion, it should lead to the
increase of the vacuum noise of measurement over the level (5).

Let us briefly summarize the results. First of all, it is clear
that the above results represent an extension and detailing of the
Mandel's model of the photon localization \cite{3}. It has been
shown that the description of the photon localization in the
process of detection and emission needs more adequate
consideration of the boundary conditions, leading to a violation
of the translation invariance inherent in the conventional model
of the plane waves of photons. This violation leads to the
qualitative change of the structure of the electromagnetic vacuum
state. In particular, the zero-point oscillations are concentrated
in vicinity of atoms, molecules, photodetectors and other local
objects which are able to interact with photons. The level of the
zero-point oscillations in vicinity of the emitting and measuring
devices can strongly exceeds that calculated as though the field
consists of the plane waves of photons. This leads to a
deterioration in the estimation of the quantum limit of precision
of measurement.

The above results can be important for different quantum optical
measurements especially in the engineered entanglement based on
the trapped atoms \cite{9}, in the experiments with atomic beams
and single-atom lasers \cite{11} and in the quantum polarization
measurement \cite{12}.

\end{document}